\documentclass[preprint,showpacs,preprintnumbers,amsmath,amssymb]{revtex4}
 \usepackage{graphicx}
 \usepackage{dcolumn}
 \usepackage{bm}
 \usepackage{ulem}
\usepackage{tensor}

\begin{document}

\title{The decline rate of the production cross-section of superheavy nuclei with $Z=114-117$ at high excitation energies}

\author{J. Hong,}
\affiliation{Department of Physics and Institute of Physics and Applied Physics, Yonsei University, Seoul 03722, Korea}

\author{G. G. Adamian,}
\affiliation{Joint Institute for Nuclear Research,
Dubna  141980, Russia}

\author{N. V. Antonenko,}
\affiliation{Joint Institute for Nuclear Research,
Dubna  141980, Russia\\
Tomsk Polytechnic University, 634050 Tomsk, Russia}

\author{P. Jachimowicz,}
\affiliation{Institute of Physics,
University of Zielona G\'{o}ra, Szafrana 4a, 65516 Zielona
G\'{o}ra, Poland}

\author{M. Kowal}
\affiliation{National Centre for Nuclear Research, Pasteura 7, 02-093 Warsaw, Poland}

\date{\today}

\begin{abstract}

The production cross sections of superheavy nuclei with charge numbers
$114-117$ are predicted in the $(5-9)n$-evaporation channels
of the $^{48}$Ca-induced complete fusion reactions for future experiments.
The estimates of synthesis capabilities are based on a uniform and consistent set of input nuclear data
provided by the multidimensional macroscopic-microscopic approach.
The contributions of various factors to the final production cross section are discussed.
As shown, the specific interplay between survival and fusion probabilities unexpectedly leads to
a relatively slow decline of the total cross-sections with increasing excitation energy.
This effect is supported by a favorable arrangement of fission barriers protecting the compound nucleus
against splitting concerning energetic thresholds for the emission of successive neutrons.
In particular, the probabilities of the formation of superheavy nuclei in the $5n$-, $6n$-,
and in some cases even in $7n$-evaporation channels are still promising.
This may offer a new opportunity for the future synthesis of unknown neutron-deficient superheavy  isotopes.

\end{abstract}

\maketitle


The question of what is the largest possible atomic number of a chemical element
in the Periodic Table is still unsettled \cite{Og1,Og1n,HofMun,Morita}.
 Currently, other unsolved problems are how to extend the chart of nuclides and
 how to explain the abundance distribution of elements in the Universe and in the Solar System \cite{Ani}.
The complete fusion reactions with  $^{48}$Ca beams and actinide targets have been successfully used to synthesize superheavy nuclei
(SHN) Cn, Nh, Fl, Mc, Lv, Ts, Og with charge numbers $Z=112-118$ in
the neutron evaporation channels ($xn$-evaporation channels, where $x$  is the number of neutrons emitted)
\cite{Og1,Og1n,Og2,SH1,SH2,SH3,SH4}
and allowed for slight approach to the so-called "the island of stability" of SHN  \cite{Og1,Og1n,HofMun}.
Most of these SHN has been obtained in the  $3n$- and $4n$-evaporation channels.
Only in the reactions $^{48}$Ca+$^{242}$Pu,$^{243}$Am,$^{245}$Cm
the evaporation residues have been detected in the $2n$-evaporation channel. The nuclei $^{285,287}$Fl
have been also produced in the $5n$-evaporation  channel.
During the transition from the $4n$-evaporation  channel to the $5n$-evaporation channel, the cross-section
dropped from about $(4-10)$ pb to about $(0.6-1)$ pb \cite{Og1n}.
However, the question of how rapidly the evaporation
residue cross section decreases with increasing beam energy is still open.
In the present paper, we want to answer this question, bearing in mind the possibilities of already
expanded/improved devices and soon possible experiments.
 Note that in our previous letter \cite{PLB2020-2}, the  production cross sections of the SHN with charge numbers
$Z=112-118$ were quite well described in $xn$-evaporation channels ($x=2-5$)
using the predictions of SHN properties from Refs. \cite{MKowal,Jach2017}.
The present article is then a natural continuation of these studies.
Employing the same mass table of Refs.~\cite{MKowal,Jach2017} based on the microscopic-macroscopic (MM) method,
 we are going to predict the chances of producing new SHN
 in the $(5-9)n$-evaporation channels
of the $^{48}$Ca-induced complete fusion reactions allowing much higher excitation energies.
Such estimates and analyzes of the corresponding excitation functions, to the best of our knowledge, are unknown in the literature.


The evaporation residue cross section
is factorized
into three independent ingredients \cite{PLB2020-2,AA,lecture,paper1}:
\begin{equation}
\sigma_{s}(E_{\rm c.m.}) =\sum_{J} \nonumber
\sigma_{cap}(E_{\rm c.m.},J)P_{CN}(E_{\rm c.m.},J)W_{s}(E_{\rm c.m.},J).  \nonumber
\label{ER_eq}
\end{equation}
In the evaporation channel, $s$ depends on the partial
capture cross section $\sigma_{cap}$
for the transition of the colliding nuclei over the entrance
(Coulomb) barrier, the probability of CN
formation $P_{CN}$ after capture and the survival probability
$W_{s}$ of excited CN which estimates the competition between fission, neutron, and charged particles evaporation
in the excited CN.
The formation of CN is calculated within  the dinuclear system model in version as described in Refs. \cite{PLB2020-2,paper1}.
This is well tested model with strong predictive power.
In Eq. (\ref{ER_eq}) the contributing angular momentum range is limited by $W_s$ and $P_{CN}$.
In the case of highly fissile SHN,  $W_s$ is a rather narrow function function of $J$ different from zero in
the vicinity of $J=0$ for all bombarding energies $E_{\rm c.m.}$.

\begin{figure}[h]
\vspace{-0.8cm}
\hspace{-0.3cm}
\includegraphics[width=0.6\textwidth, clip]{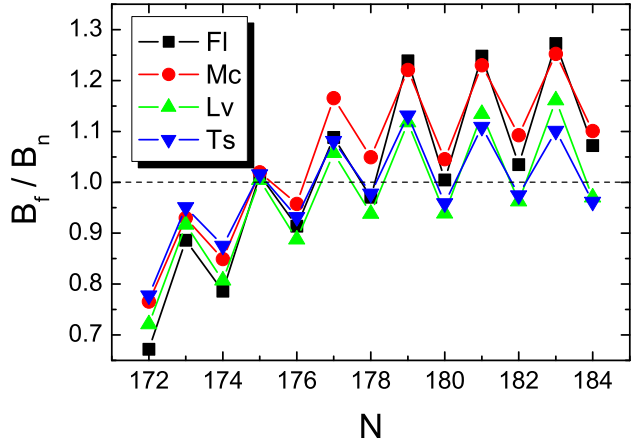}
\caption{The ratio  $B_{f}/B_{n}$ vs the neutron number $N=A-Z$ for the SHN $_{114}$Fl, $_{115}$Mc, $_{116}$Lv, $_{117}$Ts.}
\label{BfBn}
\end{figure}

\begin{figure*}[h]
\includegraphics[width=0.45\textwidth,clip]{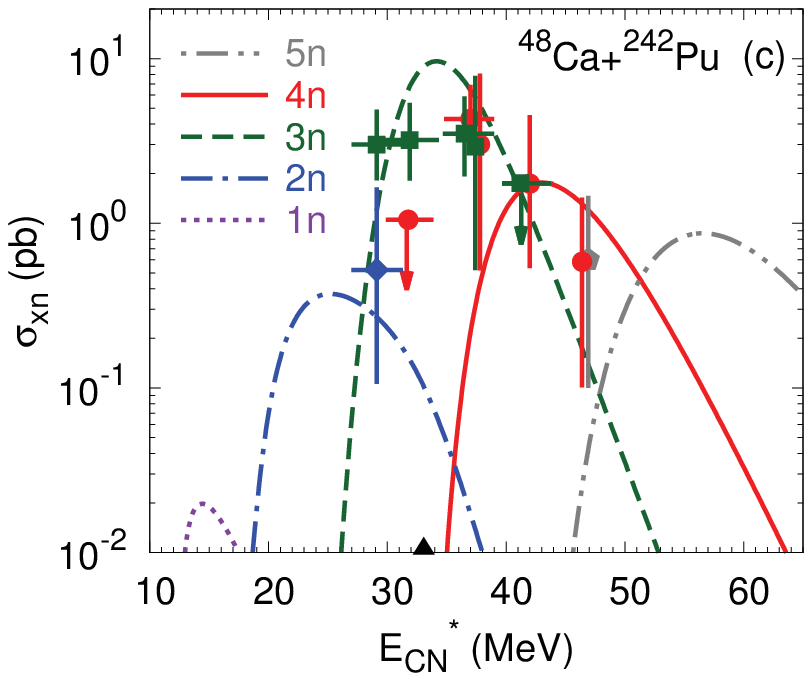}
\includegraphics[width=0.46\textwidth,clip]{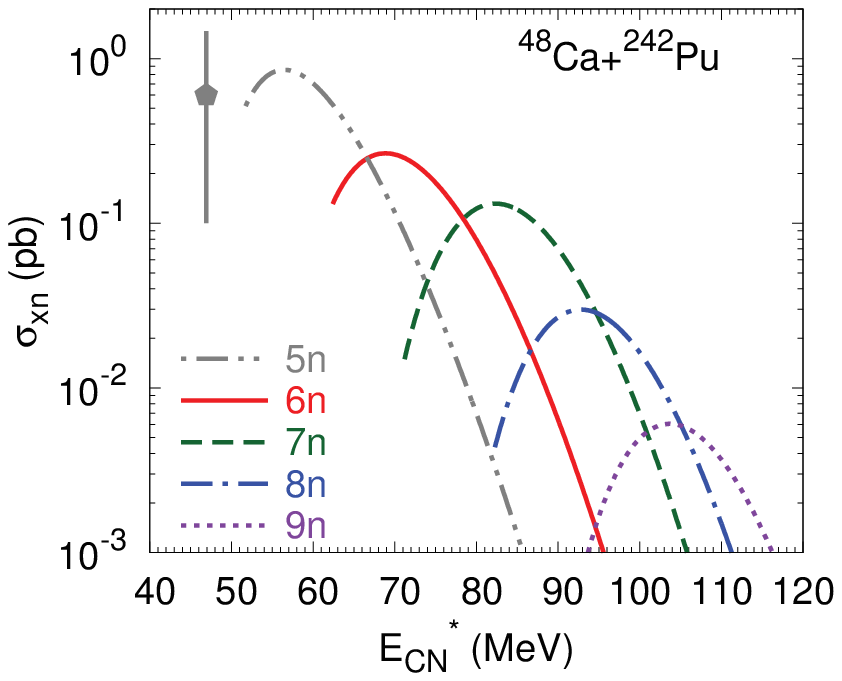}
\includegraphics[width=0.45\textwidth,clip]{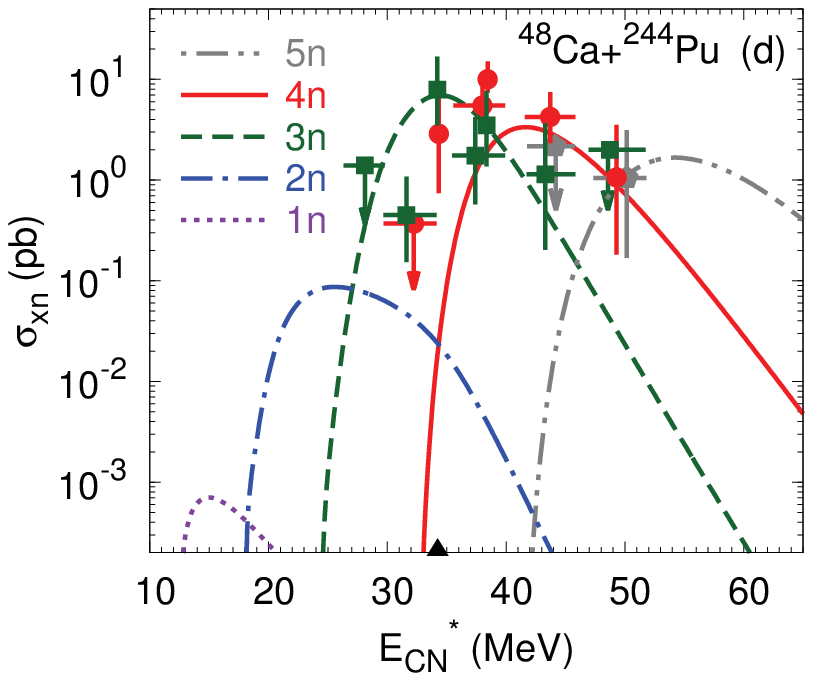}
\includegraphics[width=0.46\textwidth,clip]{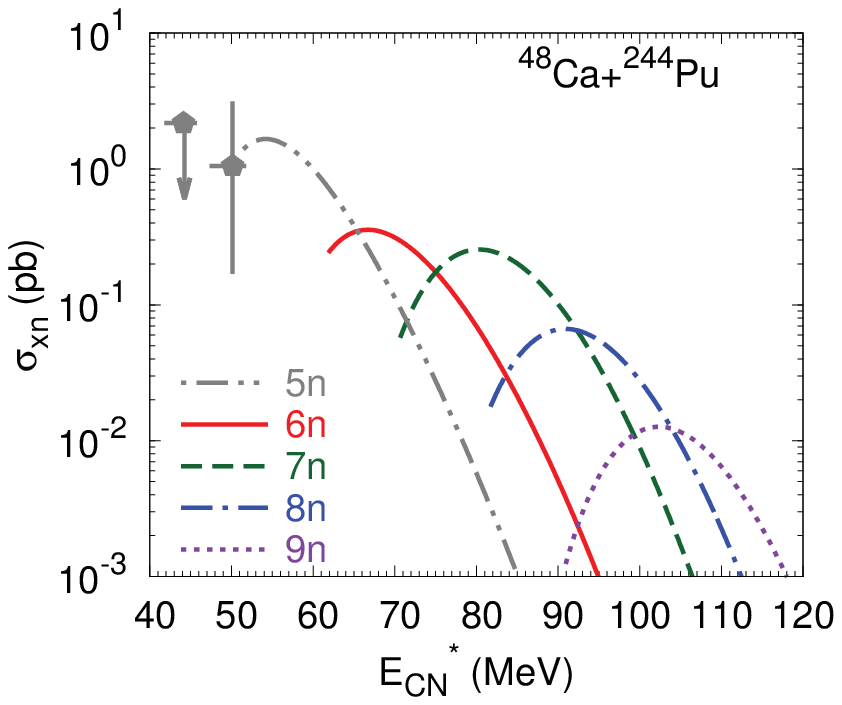}
\caption{
The  calculated (lines) excitation functions for
$xn$-evaporation channels ($x=1-9$) of the indicated complete fusion  reactions.
The mass table of Ref.~\cite{MKowal} is used in the calculations.
The mass numbers of isotopes produced are indicated.
The  black  triangles at the energy axis indicate the excitation energy
$E_{CN}^*$
of the CN at bombarding energy corresponding to the Coulomb barrier
for the sphere-side orientation.
The blue diamonds, green squares,  red circles,
and gray pentagons
represent the experimental data \cite{Og1n} with error bars for $2n$-,
$3n$-, $4n$-, and $5n$-evaporation  channels, respectively.
The symbols with the arrow indicate  the upper limits of evaporation residue cross sections.
}
\label{caact1}
\end{figure*}

\begin{figure}[h]
\includegraphics[width=0.45\textwidth,clip]{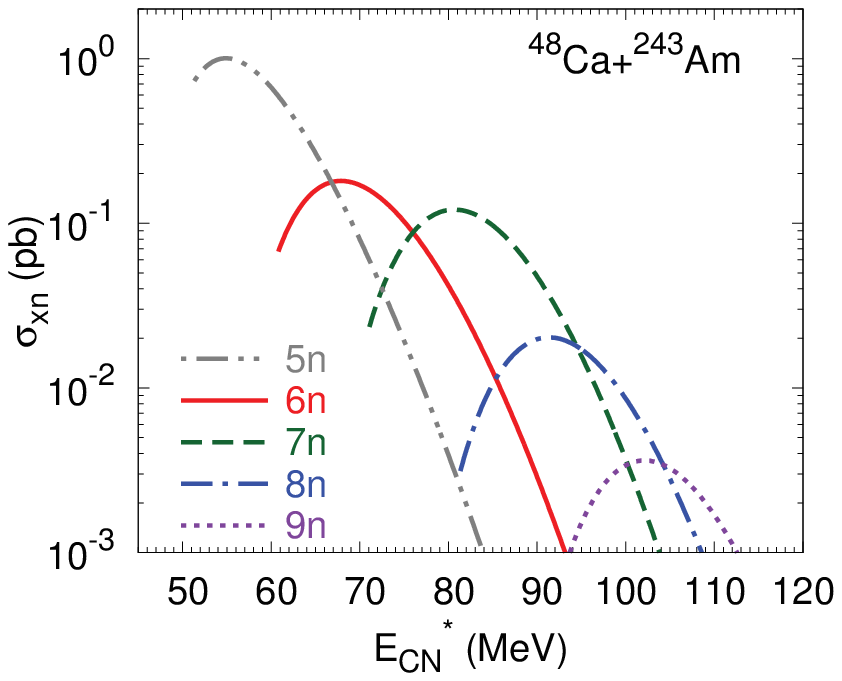}
\includegraphics[width=0.45\textwidth,clip]{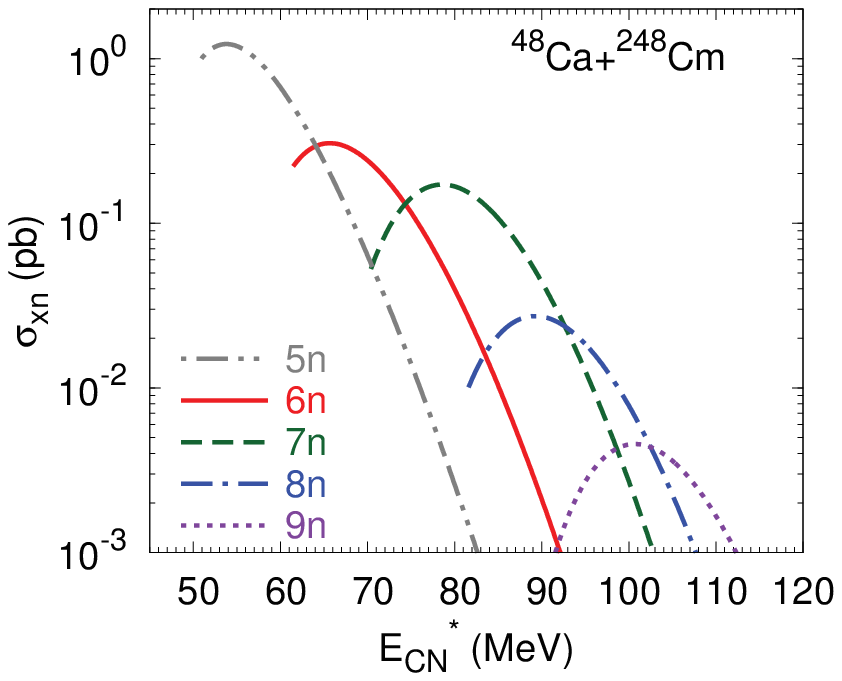}
\includegraphics[width=0.45\textwidth,clip]{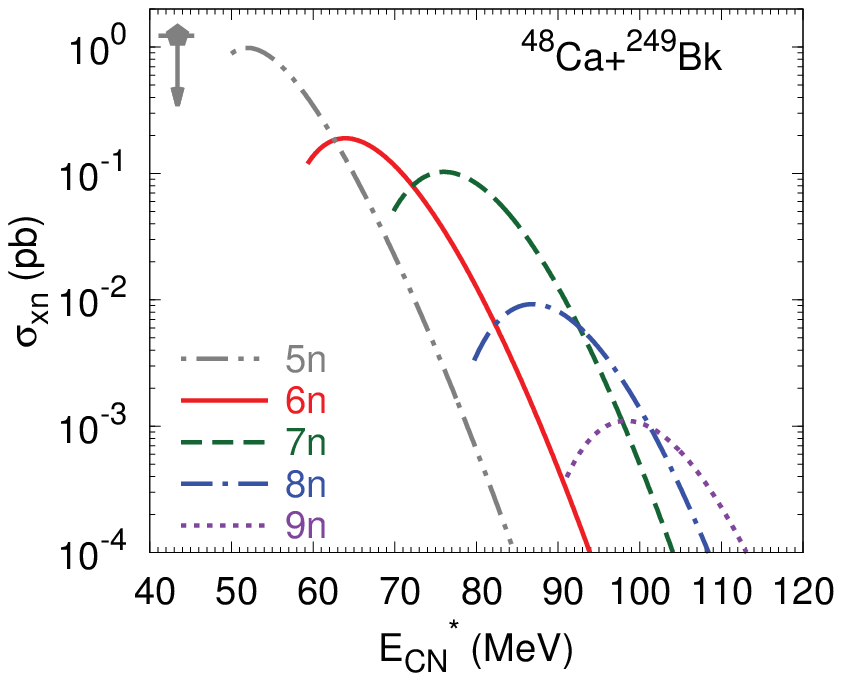}
\caption{
The same as in Fig.~\ref{caact1}, but for
$xn$-evaporation channels ($x=5-9$) of the indicated complete fusion  reactions.
}
\label{caact2}
\end{figure}

Nuclear properties required for the correct estimation
of the survival probability ($W_{s}$) were systematically calculated within the
multidimensional MM approach \cite{MKowal}.
State-of-the-art methods were used: minimization over many deformation parameters for minima and
the imaginary water flow on many-deformation energy grids for saddles, including non-axial and reflection-asymmetric shapes.
 Our  systematic calculations include odd-$A$ and odd-odd nuclei
with inner and outer fission barriers what is quite scarce in the literature.
For nuclei with odd numbers of protons, neutrons, or both, we use a standard BCS method with blocking.
One should  emphasize that the MM method used here offers
 very good agreement with various existing experimental data (nuclear masses, decay energies, fission barriers, etc.)
 in a wide range of  heaviest nuclei \cite{MKowal}.
\begin{figure}[h]
\vspace{-0.0cm}
\hspace{-0.0cm}
\includegraphics[width=0.35\textwidth,angle = -90, clip]{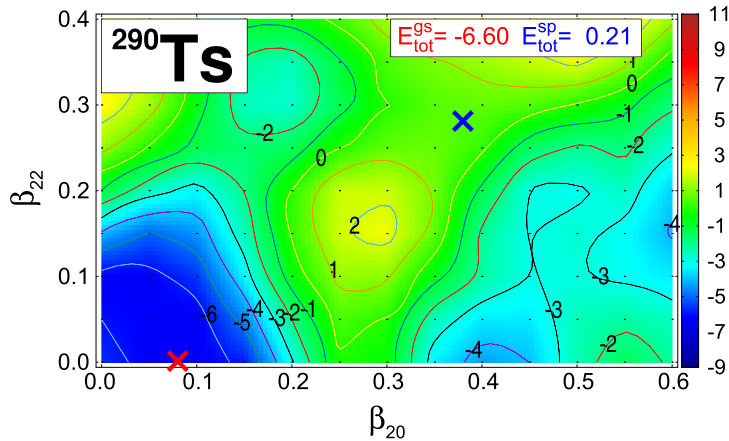}
\caption{The Potential Energy Surface (PES), $E-E_{mac}({\rm sphere})$, for $^{290}_{117}$Ts.
The crosses indicate  the saddle point (blue) and the ground state (red).}
\label{290Ts}
\end{figure}
Decisive for the survival of newly created compound nucleus is a competition between fission and neutron emission process
in the successive steps of the decay cascade.
To put it simply, the resolution of this competition is determined by the energy thresholds -
the fission barrier  $B_{f}$ and the neutron separation  energy $B_{n}$.
An example of the potential energy surface allow to find these key values
is shown in the Fig. \ref{290Ts} for the odd-odd $^{290}_{117}$Ts nucleus
while energetic relation for decay thresholds is shown in Fig. \ref{BfBn}.
A value of $B_{f}/B_{n}$  greater than one just means that the nucleus
is more protected against fission compared to the neutron emission.
This is explicitly shown in Table \ref{tab:BfBn} where the difference  $B_{f} - B_{n}$ - decisive for the
survival probability of the newly formed SHN was calculated based on the new tables of masses and fission barriers \cite{MKowal}.
Values greater than zero give (in the first order) an optimistic prediction of the new synthesis.
\begin{table}[!htbp]
  \caption{Calculated $B_{f}-B_{n}$ based on Ref. \cite{MKowal}.}
  \label{tab:BfBn}
  \centering
\begin{tabular}{|c|c|c|c|c|}
\hline
N      &   Fl   &    Mc   &   Lv   &   Ts \\
\hline
172	& 	-2.36	& -1.70	& -2.11 &-1.70   \\
173	&	-0.74	& -0.47	& -0.56	& -0.35 \\
174	& 	-1.50	& -1.07	& -1.42	& -0.92 \\
175	& 	0.09	& 0.13	& 0.03	& 0.11  \\
176	&  -0.57	&-0.29	&-0.80	&-0.48   \\
177	&	0.54	&1.06	&0.37	&0.55    \\
178	&   -0.20	&0.32	&-0.42	&-0.15   \\
179	&	1.30    &1.28	&0.70	&0.80    \\
180	&	0.03	&0.29	&-0.41	&-0.27   \\
181	&	1.32	&1.31	&0.77	&0.66    \\
182	&	0.21	&0.55	&-0.24	&-0.16   \\
183	&	1.41	&1.36	&0.88	&0.58    \\
184	&	0.41	&0.57	&-0.18	&-0.23   \\
\hline
\end{tabular}
\label{barriers}
\end{table}
The calculated excitation functions for $(5-9)n$-evaporation channels are presented in Figs.~\ref{caact1} and \ref{caact2}
for the complete fusion reactions $^{48}$Ca+$^{242,244}$Pu,$^{243}$Am,$^{248}$Cm,$^{249}$Bk. As one can see, the production cross sections
for the $(2-5)n$-evaporation channel are in quite a good agreement with the available experimental data.
As mainly seen, the rather   weak drop of the cross section with increasing excitation energy $E^*_{CN}$ is due to the interplay between the  fusion
$P_{CN}$ and the  survival $W_{xn}$ probabilities, and  due to a weak change of the difference between the fission barrier height
and neutron separation energy at $(5-9)$ steps of neutron evaporation as shown in Fig. \ref{BfBn} and in Table \ref{tab:BfBn}.
 Indeed, because  the attenuation of the shell correction at high excitation energies, the fission barrier heights become  close in $xn$- and  $(x+1)n$-evaporation channels.
%
The compliance of the excitation functions obtained under the current approach
for other target-projectile combinations leading to the already known SHN can be checked in Ref. \cite{PLB2020-2}.
Predictions for the production cross sections in high neutron emission channels are shown in the Fig. \ref{caact2} for Mc, Lv, and Ts.
One can see that according to our estimations, the production of   unknown isotopes $^{291}$Ts and $^{290}$Ts in the $6n$- and
$7n$-evaporation channels, respectively, is still quite probable as the maxima of the cross sections are of the order of tenths of a picobarns -
$\sigma^{max}_{6n}\approx 0.2$ pb and $\sigma^{max}_{7n}\approx 0.1$ pb.
Similarly optimistic are channels: $7n$ in the reactions $^{48}$Ca+$^{248}$Cm to produce the unknown isotope $^{289}$Lv or $(5-7)n$ in reaction $^{48}$Ca+$^{243}$Am which
may still quite likely leads to new isotopes $^{286}$Mc ($\sigma^{max}_{5n}\approx 1$ pb), $^{285}$Mc ($\sigma^{max}_{6n}\approx 0.2$ pb) or
to $^{284}$Mc ($ \sigma^{max}_{7n}\approx 0.1$ pb). Note that for the production of unknown neutron-deficient isotope $^{283}$Fl,
 the hot fusion reaction $^{48}$Ca+$^{242}$Pu$\to ^{283}$Fl+$7n$ ($ \sigma^{max}_{7n}\approx 0.1$ pb) looks superior to the cold fusion reaction
$^{76}$Ge+$^{208}$Pb$\to ^{283}$Fl+$1n$ for which  $\sigma^{max}_{1n}\lesssim \sigma^{exp}_{1n}$($^{70}$Zn+$^{209}$Bi)$\approx 0.02$ pb \cite{Morita}.

\begin{figure}[h]
\includegraphics[width=0.45\textwidth,clip]{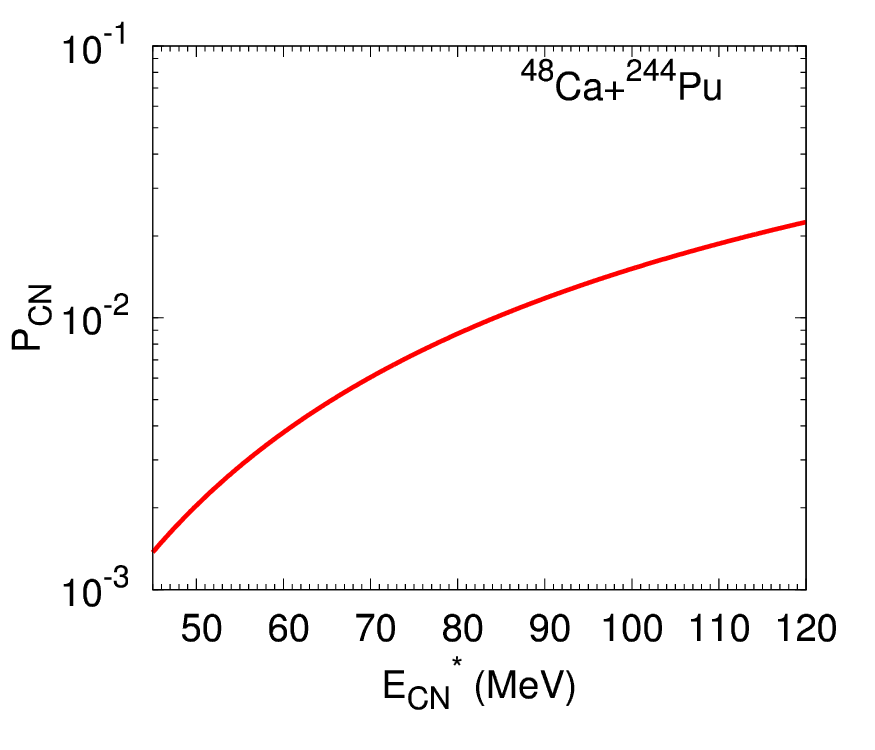}
\includegraphics[width=0.45\textwidth,clip]{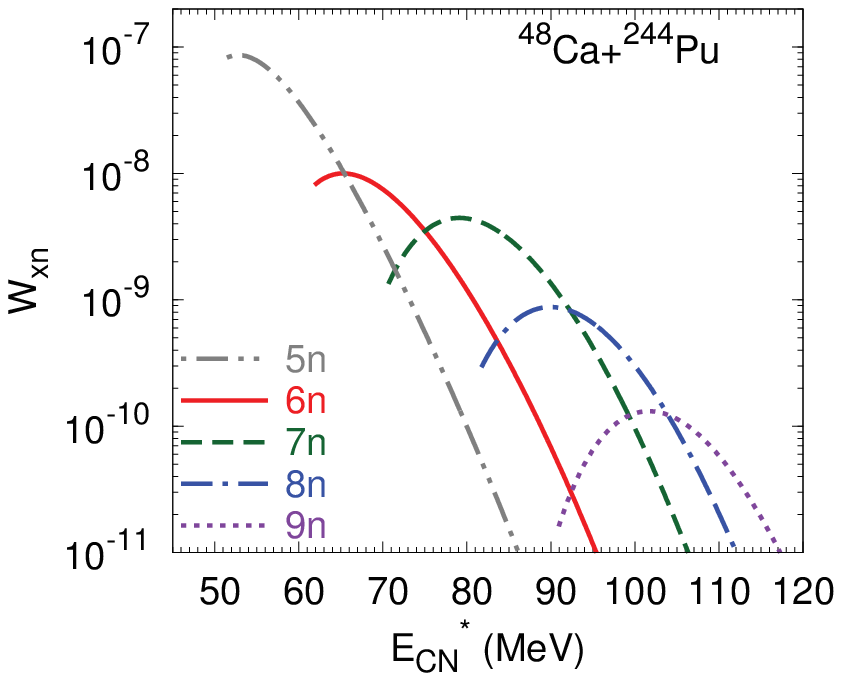}
\caption{
The calculated fusion $P_{CN}$ and survival $W_{xn}$ ($x=5-9$)
probabilities as a function of the excitation energy  for the $^{48}$Ca+$^{244}$Pu  complete fusion  reaction.
}
\label{caact3}
\end{figure}

\begin{figure}[h]
\includegraphics[width=0.45\textwidth,clip]{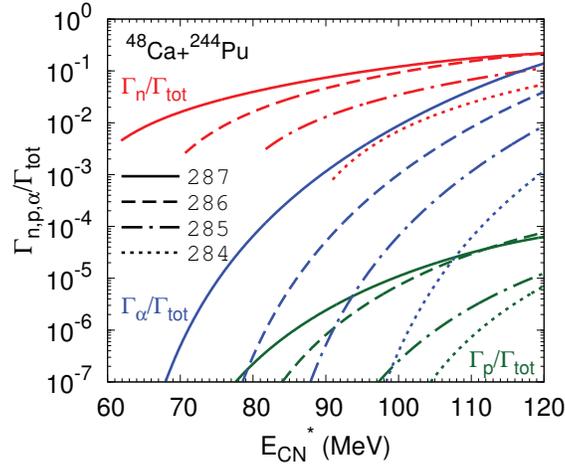}
\caption{
The calculated ratios of the widths of neutron, proton, $\alpha$-particle emissions  to the total width
as a function of excitation energy for the nuclei $^{284-287}$Fl.
}
\label{caact5}
\end{figure}

As an example, Fig. \ref{caact3} shows $P_{CN}$ and  $W_{xn}$ for the $^{48}$Ca+$^{244}$Pu  reaction.
With increasing excitation energy from 55 MeV to 102 MeV, the value of $P_{CN}$ increases by about 6 times,
while the value of $W_{xn}$ drops down by about 640 times and as a result, the production cross section decreased by more than 2 orders of magnitude.
In the case of the  $^{48}$Ca+$^{249}$Bk reaction, the fusion probability grows by about 13 times, while the survival probability decreases by about $10^{4}$.
It should be also noted that the capture cross section changes slightly in this case due to the limitation of the upper limit of angular momentum and that we
observe a similar pattern for the other reactions in Figs.~\ref{caact1} and \ref{caact2}.
%
%
%
%
%
A new fascinating possibility, we have found, for the synthesis of new superheavy isotopes
is related to alpha particles emission at the very beginning of  the entire cascade process.
Figure \ref{caact5} shows the ratio of the widths $\Gamma_{n,p,\alpha}$
in the neutron, proton, and $\alpha$-particle emission channels to the total width $\Gamma_{tot}$
as a function of the excitation energy for the different Fl isotopes. As seen, the emission
widths of proton and $\alpha$-particle grow faster with increasing excitation energy than the neutron emission width.
At energies of the order and higher than 100 MeV, the emission widths of the neutron and $\alpha$-particle  become of the same order,
which leads to the suppression of the formation of nuclei in the neutron evaporation channel.
The excitation functions of the $\alpha x'n$-evaporation channels overlap with those from $xn$-evaporation channels.


%
%
\begin{figure}[h]
\includegraphics[width=0.45\textwidth,clip]{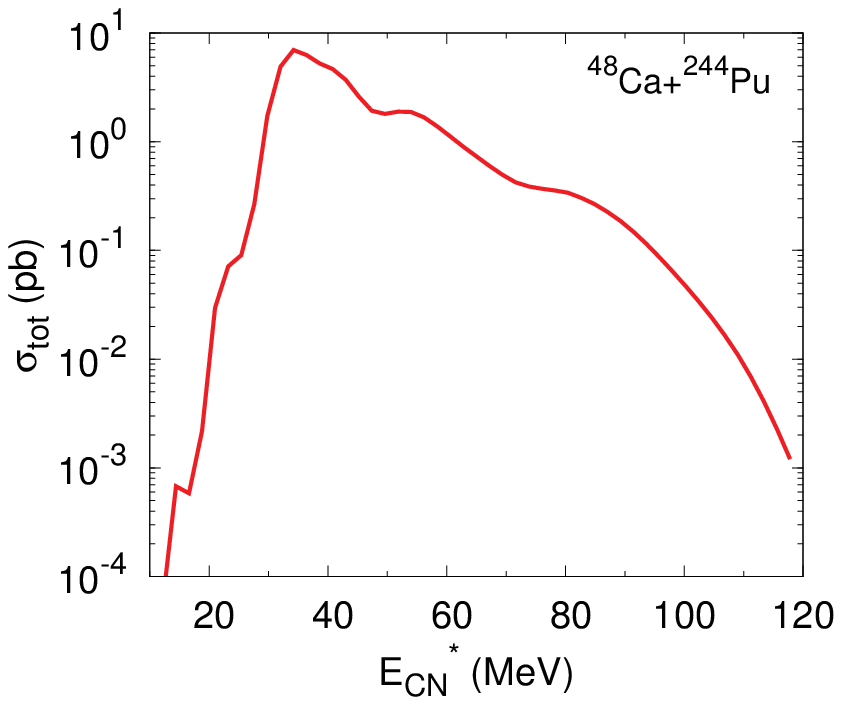}
\caption{
The calculated total cross section
as a function of the excitation energy for the indicated reaction.
}
\label{caact7}
\end{figure}
For the $^{48}$Ca+$^{244}$Pu reaction, we also present in Fig. \ref{caact7}
the total cross section
$$\sigma_{tot}(E_{\rm c.m.})=\sum_{x}\sigma_{xn}(E_{\rm c.m.}),$$
which is the sum of the cross sections of all neutron evaporation channels.
With increasing excitation energy,  from $E^*_{CN}\approx 12$ MeV, the total cross section sharply increases  by 5 orders of magnitude,
reaches a maximum at   $E^*_{CN}\approx 35$ MeV,  and then decreases relatively slowly at the $E^*_{CN}$ range of about $35-110$ MeV.

%

In conclusion, the excitation functions for the production of the new SHN with charge numbers
$Z=114-117$ were calculated within dinuclear system model  \cite{PLB2020-2,paper1}, having strong predictive power,
in $(5-9)n$-evaporation channels for the complete fusion reactions $^{48}$Ca+$^{242,244}$Pu,$^{243}$Am,$^{248}$Cm,$^{249}$Bk.
A very important element of these predictions is that they were made
based on a uniform, consistent, and systematic set of input data.
Predictions of global nuclear properties were done
by using a well-tested  multidimensional MM approach \cite{MKowal,Jach2017}.
In the presented letter, for the first time, we have indicated the possibility
of producing new isotopes of the SHN in channels with high neutron multiplicities.
 As shown, the cross section drops down from about $1$ pb to about $(1-10)$ fb at the transition from the $5n$-  to the $9n$-evaporation channel.
Thus, the decline of the cross section with increasing excitation energy unexpectedly turned out to be relatively weak.
This intriguing behavior may open up a new window for the study and the production of new isotopes of the SHN at high excitation energies.
It would be interesting to compare the production cross sections in the hot and cold fusion reactions leading to the same neutron-deficient residual
nuclei.
With the planned increase of the beam intensity (of
$\sim 10$ p$\mu$A on target) \cite{Gulbekian19}  combined with the new Dubna gas-filled magnetic recoil separator (DGFS) setup
\cite{Og1n}, such extremely low cross sections will soon be measurable.\\



\acknowledgments
We are thankful to Prof. Yu.Ts. Oganessian for the initialization of this work and fruitful discussions.
The work of G.G.A. and N.V.A. was   supported by the Ministry of Science and Higher Education of the Russian Federation
(contract 075-10-2020-117).
M.K. was co-financed by the National Science Centre under
Contract No. UMO-2013/08/M/ST2/00257 (LEA-COPIGAL).

\end{document}